\documentclass[prl,twocolumn,
 reprint, %linenumbers,
 amsmath,amssymb,
 aps, physrev,
]{revtex4-2}

\usepackage{graphicx}% Include figure files
\usepackage{dcolumn}% Align table columns on decimal point
\usepackage{bm}% bold math
\newcommand{\degree}{$^{\circ}\,$}

\newcommand{\myvec}{\textbf}

\begin{document}

%\preprint{APS/123-QED}

\title{\textbf{Pump with broadband probe experiments for single-shot measurements of plasma conditions and crossed-beam energy transfer} 
}% 
% Alt.: Pump-broadband probe experiments for single-shot measurements of plasma conditions and crossed-beam energy transfer

\author{A. Longman}
 \email{longman1@llnl.gov}
\author{R. Muir}
\author{D. E. Mittelberger}
\author{E. S. Grace}
%\author{J. Ludwig}
\author{C. Goyon}
\author{G. F. Swadling}
\author{G. E. Kemp}
\author{T. Chapman}
\author{S. Maricle}
\author{N. Vanartsdalen}
\author{A. Linder}
\author{T. Dumbacher}
\author{K. Zoromski}
\author{B. C. Stuart}
\author{F. Albert}
\author{J. E. Heebner}
\author{P. Michel}
 \email{michel7@llnl.gov}

\affiliation{Lawrence Livermore National Laboratory, 7000 East Ave, Livermore CA 94550, USA}%

\date{\today}

\begin{abstract}
A novel technique for measuring plasma conditions using monochromatic pump-broadband probe laser interactions has been experimentally demonstrated. Originally proposed in [J. Ludwig et al., Phys. Plasmas \textbf{26}, 113108 (2019)], this method utilizes crossed-beam energy transfer between the broadband probe and the pump, mediated by plasma ion acoustic waves. The complete energy transfer spectrum can be captured in a single shot, enabling the inference of plasma parameters such as density, electron and ion temperatures, and flow velocity. Compared to Thomson scattering, this technique offers signal enhancements typically larger than 9 orders of magnitude, significantly reducing the required probe laser intensity and facilitating interactions that are linear and measurements that are non-perturbative of the plasma. Furthermore, it provides a powerful tool for advancing studies of crossed-beam energy transfer under conditions relevant to inertial confinement fusion experiments.
%Crossed-beam energy transfer between a broadband probe and an intense, monochromatic pump laser was experimentally demonstrated as a novel way to diagnose plasma conditions. 

%An article usually includes an abstract, a concise summary of the work
%covered at length in the main body of the article. 

\end{abstract}

\maketitle

Thomson scattering (TS) has long been an essential method for diagnosing plasma conditions. In high-energy density (HED) or inertial confinement fusion (ICF) experiments, TS typically utilizes a probe laser beam that scatters off plasma fluctuations within the collective regime \cite{FroulaBook}. At a fixed collection angle relative to the probe's propagation direction, the scattered light spectrum reveals distinct peaks associated with electron and ion plasma modes. These features enable the determination of key plasma parameters, including density, electron and ion temperatures, flow velocity, and even the electron distribution function \cite{TurnbullNP20}.

One of the primary challenges of TS is its low cross-section. The ratio of scattered power to probe power, integrated over all frequencies, is expressed as $P_s/P_i=(8\pi/3)n_er_e^2Ld\Omega$, where $n_e$ is the electron density, $r_e$ is the classical electron radius, $L$ is the length of the scattering volume along the probe direction and $d\Omega$ is the collection solid angle. For typical plasma conditions relevant to ICF or HED experiments, such as $n_e$ = 10$^{19}$ cm$^{-3}$, $L$ = 50 $\mu$m and $d\Omega$ = 0.1 (corresponding to f/8 collection optics), we have $P_s/P_i\approx 3\times 10^{-9}$. These extremely small signal levels compete against noise emission from the target and necessitate increasing the intensity of the probe laser. However, higher probe intensity introduces additional challenges, such as nonlinear propagation effects (e.g., filamentation and self-focusing) and modification of local plasma conditions by the probe itself.

An alternative to Thomson scattering for measuring plasma conditions was proposed in Ref.~\cite{Ludwig2019}. This concept relies on using a broadband probe interacting with a pump laser in a plasma. For sufficiently high pump intensities and appropriate plasma conditions, each frequency component of the probe can couple to the pump via the crossed-beam energy transfer (CBET) process. During this interaction, the probe frequencies lose or gain energy to or from the pump when the beat wave between the pump and probe is in the vicinity of an ion acoustic wave resonance. As a result, the plasma ion mode spectrum becomes imprinted onto the post-interaction spectrum of the probe. This process is fundamentally similar to TS off ion modes, except that the measured signal is the probe beam itself, coupled to the pump via plasma waves, rather than scattered light coupled to the probe as in TS. Consequently, the measured signal level can be dramatically enhanced compared to TS, typically by more than 9 orders of magnitude. Additionally, this technique enables single-shot measurement of the full CBET spectrum at very high spectral resolution, eliminating the need for multiple shots to scan the pump–probe frequency shift, a process that introduces shot-to-shot uncertainty and is particularly challenging in ICF facilities, where available shots are limited \cite{Turnbull2017,GlenzerScience10,TurnbullNP20}.

\begin{figure*}
\includegraphics{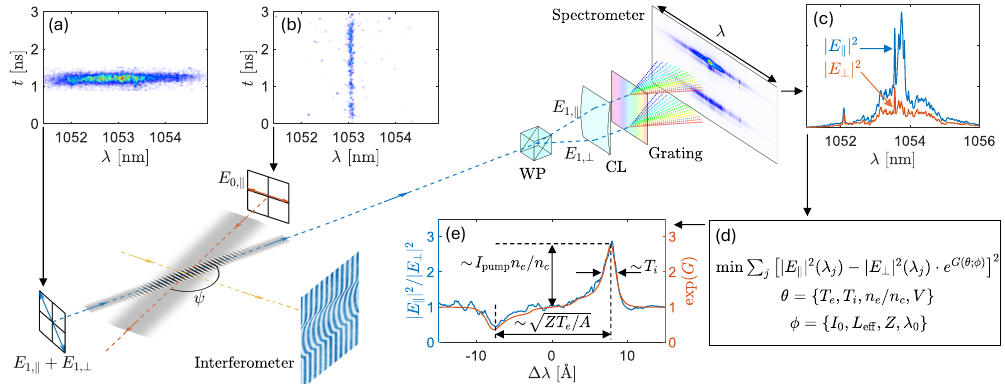}% Here is how to import EPS art
\caption{Schematic of the broadband CBET experiment. The horizontally polarized pump ($E_0$,b) transfers energy to and from the broadband probe ($E_1$,a) via ion-acoustic waves, with energy exchange depending on wavelength difference $\Delta\lambda=\lambda_1-\lambda_0$, and crossing angle $\psi$. The probe's amplified (parallel) and un-amplified (perpendicular) components are separated using a Wollaston prism (WP), re-imaged with cylindrical lenses (CL), and measured by a high-resolution spectrometer (c). Nonlinear regression (d) fits a gain curve to the measured amplification extracting plasma parameters (e), and is compared to plasma densities measured by interferometry.}\label{figure1}
\end{figure*}

In this Letter, we present the first experimental demonstration of multi-wavelength CBET in the linear regime, and the concept of a pump–broadband-probe technique for single-shot experimental tests of CBET physics and plasma parameter determination. A low-energy probe beam with a fractional bandwidth of $\approx$ 0.4\% was used to measure plasma conditions in the presence of an intense pump laser. By fitting the gain curve using the same kinetic model commonly used for Thomson scattering (TS) \cite{FroulaBook} or CBET calculations \cite{Williams04,MichelPRL09}, we obtained measurements of plasma parameters, including electron and ion temperatures, flow velocity, and even density. Notably, density measurements, normally requiring the probing of electron plasma modes in TS, were successfully achieved using this technique, and validated by an independent interferometry measurement to within 14\%. This method could, in principle, be applied to experiments at ICF facilities such as the National Ignition Facility (NIF), where efforts to develop TS diagnostics have been hindered by the low signal-to-noise ratio of the TS probe, and open new avenues for single-shot studies of CBET in ICF at any other current or future laser facility.

%Resonant energy transfer between two intersecting laser beams with frequencies and wavevectors ($\omega_i,k_i$) can occur when crossed at an angle $\psi$, provided their polarization vectors are co-planar and their frequency difference is close to the ion-acoustic resonance condition, $\Delta\omega\approx\left|k_bc_s-k_b\cdot V\right|$, where $V$ is the plasma flow velocity. The beat wave-vector $k_b$ can given approximately as $2k_0\sin(\psi/2)$, and the ion-acoustic sound speed is given for a single species plasma as $c_s=\sqrt{(ZT_e+3T_i)/m_i }$, with $Z$, $T_e$ and $T_i$ as the ionization state, electron, and ion temperatures respectively, and $m_i$ the ion mass. The energy transfer rate from the pump ($i=0$) to the probe ($i=1$) is modeled by the line integrated gain: $I(s)=I(0)\exp\left[\int G(s)ds\right]$ with the gain coefficient defined as $G=-2k_1 \Im(\delta\eta)/\eta_0$ where $\delta\eta$ denotes the perturbation to the refractive index induced by the beating of the two beams, and $\eta_0$ is the unperturbed plasma refractive index \cite{Michel2014}. Energy transfer cannot occur for orthogonal polarization components

Experiments were conducted at the Jupiter Laser Facility (JLF) using the Janus laser system and STILETTO, a novel device for spectral-temporal pulse shaping \cite{Mittelberger:21,Mittelberger:22}. The horizontally polarized pump beam $E_0$ delivered $\sim$300 J in 3 ns square pulses centered at $\lambda_0=1053.01\pm0.03$ nm, as shown in Fig. \ref{figure1}b. It was focused by an $f$/6.7 lens and spatially smoothed by a 600 $\mu$m continuous phase plate (CPP), producing an average focal-plane intensity of $\sim3.2\times10^{13}$ W/cm$^2$. The pump served two roles: (1) ionizing CH$_4$ gas injected from a 3 mm gas jet to produce plasmas with electron densities of 0.1 -- 0.7$\%$ of critical and temperatures of $\sim$200 eV (electrons) and $\sim$20 eV (ions); and (2) to imprint the ion-acoustic response onto the broadband-probe beam via CBET. 

The broadband-probe $E_1$ (Fig. \ref{figure1}a), carrying 5 mJ of energy, was polarized at $\sim$45\degree relative to the pump, and focused using a $f$/6.7 lens and a 200 $\mu$m CPP. It counter-propagated at a 153\degree crossing angle to minimize $f$-number effects on the gain spectrum \cite{OudinPOP25}. A new regenerative amplifier on the Janus East beamline provided $>4$ nm of bandwidth. The probe was spectrally shaped in amplitude and phase via STILETTO, then compressed to 167 ps FWHM, allowing ample time for IAWs to develop while minimizing plasma evolution over the probe duration \cite{DivolPOP19}. The pump-probe delay was adjustable to sample evolving plasma conditions.

After amplification, the probe was collimated and transported to a Wollaston prism (WP), which split its horizontal and vertical polarization components (Fig. \ref{figure1}, WP). These diverging beams were re-imaged into a high-resolution spectrometer (resolution $<$25 pm) using cylindrical lenses (CL). Although the probe carried only 5 mJ of energy, attenuation by a factor of 10$^{-4}$ was necessary to avoid saturating the detector, highlighting the significantly stronger signal levels of this measurement compared to TS. Due to shot-to-shot spectral intensity variations, a per-shot reference was required: similar to Ref. \cite{Turnbull2017}, since only the parallel polarization is amplified, the perpendicular component served as an unamplified reference (Fig. \ref{figure1}c). Additionally, a third short-pulse 800 nm probe laser, co-timed with the broadband-probe and propagating orthogonally to the pump beam, was used to independently measure the spatial plasma density profile via interferometry.

\begin{figure}[h]
\includegraphics{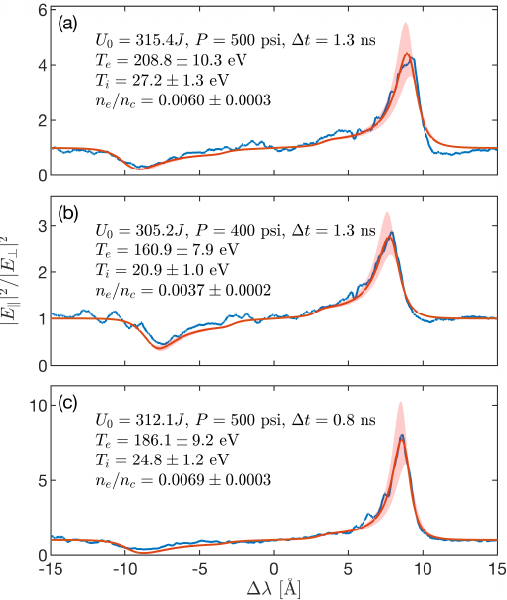}% Here is how to import EPS art
\caption{Representative CBET spectra (blue) and corresponding gain curves (red) and uncertainties (shaded). Pump laser energy $U_0$, gas jet backing pressure $P$, and pump-probe delay $\Delta t$ is given within each subplot, as well as the fitted electron temperature $T_e$, ion temperature $T_i$, and density $n_e/n_c$ with associated errors. }\label{figure2}
\end{figure}

Example gain spectra and corresponding fits for three representative shots are shown in Fig. \ref{figure2}, with fitting parameters and associated errors listed within each panel. Figures \ref{figure2}a--b were acquired with a pump-probe delay of $\Delta t=1.3$ ns at backing pressures of 500 and 400 psi, respectively, while Fig. \ref{figure2}c corresponds to $\Delta t=0.8$ ns at the same pressure as (a). Each spectrum clearly resolves the two-peak structure characteristic of CBET and also exhibits distinct multi-species features associated with fast and slow acoustic modes, visible as spectral ``shoulders'' near $\pm3$ \text{\AA} in (a) and (b) \cite{Williams95}.

To extract the gain factor $|E_{1,\parallel}|^2/|E_{1,\perp}|^2=\exp(G)$, we perform nonlinear regression \cite{Swadling22} to optimize the gain parameters $\theta(T_e,T_i,n_e/n_c,V)$ by minimizing the residual $\left[|E_{\parallel}|^2-|E_{\perp}|^2\cdot\exp(G(\theta;\phi))\right]^2$. The gain is computed using the standard linear kinetic CBET model,
\begin{equation} \label{eq1}
    G=L_{\mathrm{eff}}\frac{|\myvec{k}_0-\myvec{k}_1|^2}{4k_0^2}\text{Im}\left[ \frac{\chi_e(1+\chi_i)}{1+\chi_e+\chi_i}\right] a_0^2 \,,
\end{equation}
where $\myvec{k}_{0,1}$ is the wave-vector of the pump (0) or probe (1), $\chi_e$ and $\chi_i$ are the electron and ion susceptibilities evaluated at $[(\omega_0-\omega_1)-(\myvec{k}_0-\myvec{k}_1)\cdot \myvec{V}, \myvec{k}_0-\myvec{k}_1]$ and $a_0\approx 0.855\times 10^{-9}\sqrt{I_0 [\text{W/cm}^2]\lambda_0 [\mu\text{m}]}$ is the pump's normalized vector potential \cite{MichelBook}. Additional experimental inputs, $\phi(I_0,L_{\mathrm{eff}}, Z,\lambda_0)$ and their associated uncertainties include the ionization state $Z$, pump intensity $I_0$, and the effective gain length $L_{\mathrm{eff}}=\langle I_0\rangle^{-1}\int{I_0(s)ds}=1.11\pm0.03$ mm, estimated from far-field images \cite{Ludwig2019}. This probed volume corresponds to a cylindrical plasma of length $1.11$ mm and radius $100$ $\mathrm{\mu m}$. It was found that fitting the signal difference, rather than their ratio, reduces noise and avoids singularities when the reference signal approaches background levels. 

We fit the experimental signal in linear space without additional weighting. While fitting $\log(\mathrm{signal})$ (i.e., the gain exponent $G$) could, in principle, improve accuracy by equally weighting the antisymmetric gain peaks about $\Delta\lambda=0$, background noise degrades the negative peak and worsens the fit. Improving background levels would likely enhance fit quality and reduce plasma parameter uncertainties beyond those presented in this proof-of-principle Letter. Uncertainties of the fitted gain curves were estimated from the standard errors of the fits, uncertainties in the input parameters $(\delta I_0,\delta L_{\mathrm{eff}},\delta\lambda_0)$, and by bounding each fitted variable to assess the resulting spread in the others. The uncertainties in the fits are shown as shaded red regions in Fig.\ref{figure2}. 

Fits assumed fully ionized CH$_4$ consistent with prior experiments using a similar configuration \cite{Turnbull2017}, and informed by HYDRA simulations \cite{Marinak2001}. While improved estimates of the average ionization $\langle Z\rangle$ might help refine inferred plasma temperatures \cite{Cho25}, such modeling is beyond the scope of this Letter. The retrieved fitted values of temperature and density are given for each of the configurations in Fig. \ref{figure2}. Flow velocities determined from the fits were found to be on the order of $V/c_s\sim0.01$, in agreement with simulations, resulting in negligible spectral shifts of approximately $13$ pm, roughly half the available spectral resolution. 

The fitting of the gain curve also assumed Maxwellian distributions, but could be extended to include the Langdon effect \cite{LangdonPRL80,MattePPCF88,TurnbullNP20} or any other source of non-Maxwellian distribution features. The impact of the Langdon effect can be estimated via the factor $\alpha=Z_{eff}a_0^2c^2/v_{Te}^2$, where $Z_{eff}=\langle Z^2\rangle/\langle Z \rangle$ is the effective charge, $c$ is the speed of light and $v_{Te}\approx 1.33\times 10^7\sqrt{T_e\,[\text{keV]}}$ is the electron thermal velocity. In our conditions ($Z_{eff}=4$, $I\approx 3\times 10^{13}$ W/cm$^2$, $T_e\approx$ 200 eV), we have $\alpha\approx 0.248$, indicating that the deviation from Maxwellian should be small but measurable, with a superGaussian exponent $m=2.5$ (vs. 2 for a purely Gaussian electron distribution).

\begin{figure}[ht]
\includegraphics{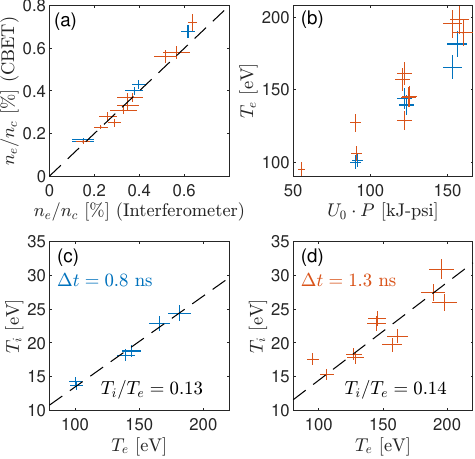}% Here is how to import EPS art
\caption{(a) Comparison of electron densities measured via cross-beam energy transfer (CBET) and interferometry; the dashed line indicates the ideal 1:1 correspondence. (b) Electron temperature as a function of the product of pump energy and gas pressure. (c), (d) Ion temperature plotted against electron temperature for delay times of $\Delta t = 0.8$ ns and $\Delta t = 1.3$ ns, respectively. In all panels, measurements at $\Delta t = 0.8$ ns are shown in blue, and those at $\Delta t = 1.3$ ns are shown in red.}\label{figure3}
\end{figure}

To validate our results, we compare the densities extracted from the fitted gain curves with those measured in situ using the interferometer (Fig. \ref{figure3}a), finding excellent agreement across the full range of values for both delay times, $\Delta t=0.8$ ns (blue) and 1.3 ns (red). The corresponding electron temperatures are shown in Fig. \ref{figure3}b as a function of the product of pump energy and gas jet backing pressure, revealing a linear trend. Ion temperatures, plotted against electron temperatures for each delay time, are shown in Figs. \ref{figure3}c and \ref{figure3}d. The average ion-to-electron temperature ratios are 0.13 and 0.14 for 0.8 ns and 1.3 ns, respectively. These measurements were compared to HYDRA simulations performed using the same laser fluence as in the experiment, with initial molecular densities adjusted to achieve an electron density of $n_e/n_c=0.0037$, consistent with the values observed in Fig. \ref{figure2}b. While the simulated electron temperatures roughly agree with experimental observations ($T_e=142$ eV, vs. 161 eV measured), the ion temperatures are significantly lower ($T_i=12.4$ eV, vs. 21 eV measured), which yields $T_i/T_e=0.056$ and 0.087 for the same delays.

The broadband probe concept could, in principle, be implemented at large-scale ICF laser facilities, where high background levels and limited shot numbers make single-shot diagnostics essential. In such facilities, the drive beams are typically narrowband 351 nm lasers, which could serve as pumps for a low-energy broadband probe intersecting them in an underdense plasma region, either within the laser-entrance hole (LEH) of a hohlraum or in the coronal plasma of a directly driven capsule. In this configuration, the broadband probe would be centered at 351 nm and require a percent-level fractional bandwidth, as recently demonstrated on the FLUX laser system \cite{Froula25}.

\begin{figure}[t]
\includegraphics[width=0.9\linewidth]{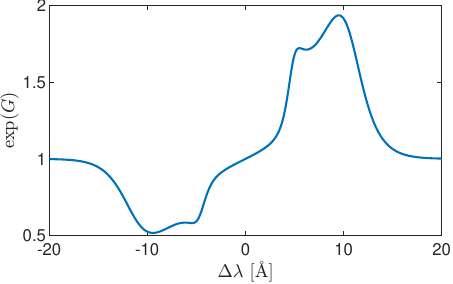}% Here is how to import EPS art
\caption{
Example amplification factor for a uniform CH\(_4\) plasma with \(\lambda_0  = 351~\mathrm{nm}\), \(n_e = 0.05\,n_c\), \(T_e = 2~\mathrm{keV}\), \(T_i = 0.5~\mathrm{keV}\), \(I_{\mathrm{pump}} = 1 \times 10^{14}~\mathrm{W/cm^2}\), \(\psi = 153^{\circ}\), and \(L_{\mathrm{eff}} = 1.3~\mathrm{mm}\).
}
\label{figure4}
\end{figure}

Figure \ref{figure4} presents the calculated probe amplification as a function of wavelength, assuming uniform density and temperature, under representative ICF plasma conditions: CH$_4$, $n_e=0.05n_c$, $T_e$ = 2 keV, $T_i$ = 0.5 keV, $I_{\mathrm{pump}}=1\times 10^{14}$ W/cm$^2$, $\psi=153^{\circ}$, and an effective gain length $L_{\mathrm{eff}}\approx w_0/ \sin(\psi)=1.3$ mm, where $w_0\approx1$ mm is the pump spot size. Compared to previous demonstrations, the electron temperature here is ten times higher, increasing the ion-acoustic speed $c_s\propto\sqrt{ZT_e}$. However, because both the pump and probe wavelengths are three times shorter, the characteristic spacing between gain peaks \cite{MichelBook}, 
\begin{equation}
    \Delta\lambda\approx \frac{2\lambda_0}{c}\sin\left(\frac{\psi}{2}\right)\sqrt{\frac{Z}{A}\frac{T_e}{m_p}}, 
\end{equation}
remains comparable, where $m_p$ and $A$ are the proton mass and atomic number, respectively. 

TS measurements at ICF facilities typically require probe energies of tens to hundreds of joules to overcome high background levels and low signal yields. A broadband probe can reduce this energy requirement by orders of magnitude. For example, a typical NIF shot produces $\approx 1\mathrm{mJ/cm^2}$ of background 351 nm light on the chamber walls (5 m radius) near the 
18$^\circ$ ports, locations that could be well suited for probe injection. To produce a probe signal several times brighter than this background, the required probe energy would be less than 1 J, depending on the aperture size of the collection optics.

In summary, we have demonstrated for the first time the use of a broadband probe to measure plasma conditions and the CBET spectrum in laboratory experiments under conditions relevant to ICF and HED physics. In contrast to Thomson scattering, this approach requires the presence of a pump laser in the plasma, with an intensity compatible with measurable gain. However, it eliminates the low-signal limitation of TS, and there is widespread interest in probing plasmas where intense lasers are already present for a broad range of ICF and HED experiments. Most of these experiments employ laser intensities that are well suited for coupling to a low-energy broadband probe. The measured plasma conditions are integrated over the overlap volume between the probe and pump beams; this integration volume, and the corresponding broadening of the gain curve, can be minimized by adjusting the beam sizes and pointing. Beyond plasma diagnostics, such a broadband probe could significantly advance the understanding and characterization of CBET at NIF and other ICF facilities by enabling measurement of the full CBET spectrum in a single shot.

This work was supported by the U.S. DOE Office of Science, Fusion Energy Sciences under Field Work Proposal No. SCW1836-1: LaserNetUS: Discovery Science and Inertial Fusion Energy research at the Jupiter Laser Facility, and Lawrence Livermore National Laboratory, under Contract No. DE-AC52-07NA27344. This work was supported in part by the Laboratory Directed Research and Development program under 24-ERD-031 and 25-LW-113. The authors would like to thank N. Lemos, J. M. Di-Nicola, J. Moody, J. Ludwig, A. Vella, and A. Oudin for useful discussions.

\bibliography{references}% Produces the bibliography via BibTeX.

\end{document}